%% file: malpaca.tex
\documentclass[graybox]{svmult}

\usepackage{type1cm}        
\usepackage{makeidx}         
\usepackage{graphicx}        
\usepackage[bottom]{footmisc}

\usepackage{newtxtext}       %
\usepackage{newtxmath}       

\usepackage{pifont}
\usepackage{makecell}
\usepackage{multirow,multicol}
\usepackage{float}
\usepackage{subfig}
\usepackage{booktabs} 

\newcommand{\nop}[1]{}

\newcommand*\OK{\ding{51}}
\def\BibTeX{{\rm B\kern-.05em{\sc i\kern-.025em b}\kern-.08em
    T\kern-.1667em\lower.7ex\hbox{E}\kern-.125emX}}

\makeindex             

\begin{document}

\title*{Beyond Labeling: Using Clustering to Build Network Behavioral Profiles of Malware Families}
\titlerunning{Using Clustering to Build Network Behavioral Profiles of Malware Families}
\author{Azqa Nadeem, Christian Hammerschmidt, Carlos H. Ga\~n\'an, Sicco Verwer}

\nop{
\authorrunning{A. Nadeem et al.}
\institute{Azqa Nadeem: \email{azqa.nadeem@tudelft.nl}, 
\and Christian Hammerschmidt: \email{c.a.hammerschmidt@tudelft.nl}, 
\and Carlos H. Ga\~n\'an: \email{c.hernandezganan@tudelft.nl},  
\and Sicco Verwer:\email{s.e.verwer@tudelft.nl}, \at Delft University of Technology, The Netherlands}}

\institute{Azqa Nadeem, \at Delft University of Technology, The Netherlands, \email{azqa.nadeem@tudelft.nl}
\and Christian Hammerschmidt, \at Delft University of Technology, The Netherlands, \email{c.a.hammerschmidt@tudelft.nl}
\and Carlos H. Ga\~n\'an, \at Delft University of Technology, The Netherlands, \email{c.hernandezganan@tudelft.nl}
\and Sicco Verwer, \at Delft University of Technology, The Netherlands,  \email{s.e.verwer@tudelft.nl}}
\authorrunning{A. Nadeem, C. Hammerschmidt, C. H. Ga\~n\'an, S. Verwer }
%
%
\maketitle

\abstract*{Malware family labels are known to be inconsistent. They are also black-box since they do not represent the capabilities of malware. The current state-of-the-art in malware capability assessment include mostly manual approaches, which are infeasible due to the ever-increasing volume of discovered malware samples. \\
We propose a novel unsupervised machine learning-based method called MalPaCA, which automates capability assessment by clustering the temporal behavior in malware's network traces. MalPaCA provides meaningful behavioral clusters using only 20 packet headers. Behavioral profiles are generated based on the cluster membership of malware's network traces. A Directed Acyclic Graph shows the relationship between malwares according to their overlapping behaviors.
The behavioral profiles together with the DAG provide more insightful characterization of malware than current family designations. 
We also propose a visualization-based evaluation method for the obtained clusters to assist practitioners in understanding the clustering results.
We apply MalPaCA on a financial malware dataset collected in the wild that comprises of 1.1k malware samples resulting in 3.6M packets. 
Our experiments show that (i) MalPaCA successfully identifies capabilities, such as port scans and reuse of Command and Control servers; (ii) It uncovers multiple discrepancies between behavioral clusters and malware family labels; and (iii) It demonstrates the effectiveness of clustering traces using temporal features by producing an error rate of 8.3\%, compared to 57.5\% obtained from statistical features. }

\abstract{Malware family labels are known to be inconsistent. They are also black-box since they do not represent the capabilities of malware. The current state-of-the-art in malware capability assessment include mostly manual approaches, which are infeasible due to the ever-increasing volume of discovered malware samples. 
We propose a novel unsupervised machine learning-based method called MalPaCA, which automates capability assessment by clustering the temporal behavior in malware's network traces. MalPaCA provides meaningful behavioral clusters using only 20 packet headers. Behavioral profiles are generated based on the cluster membership of malware's network traces. A Directed Acyclic Graph shows the relationship between malwares according to their overlapping behaviors.
The behavioral profiles together with the DAG provide more insightful characterization of malware than current family designations. 
We also propose a visualization-based evaluation method for the obtained clusters to assist practitioners in understanding the clustering results.
We apply MalPaCA on a financial malware dataset collected in the wild that comprises of 1.1k malware samples resulting in 3.6M packets. 
Our experiments show that (i) MalPaCA successfully identifies capabilities, such as port scans and reuse of Command and Control servers; (ii) It uncovers multiple discrepancies between behavioral clusters and malware family labels; and (iii) It demonstrates the effectiveness of clustering traces using temporal features by producing an error rate of 8.3\%, compared to 57.5\% obtained from statistical features. }

\input{Sec-Intro}

\input{Sec-Litrev}

\input{Sec-Method}

\input{Sec-Results}

\input{Sec-Limitations}

\input{Sec-Conclusions}

\bibliographystyle{plain}
\bibliography{malpaca}

\end{document}

%% file: Sec-Intro.tex
\section{\textbf{Introduction}}\label{sec:intro}
The first malware was discovered over thirty years ago. Yet, it is still one of the leading threats in cybersecurity\footnote{https://www.cybersecurity-insiders.com/top-15-cyber-threats-for-2019/}. AV-test, a security research institute, reported detecting over 1000 Million malware samples in 2019\footnote{https://www.av-test.org/en/statistics/malware/}. 
Anti-Virus (AV) companies play a pivotal role in analyzing malware by assigning labels to newly discovered samples. 
However, there are several shortcomings of malware family labels:
(i) Each vendor has its  own way of determining a malware family. Labels obtained from different vendors are often inconsistent~\cite{kantchelian2015better}. 
(ii) The precise methods used by each vendor are proprietary and unstandardized~\cite{sebastian2016avclass}.
(iii) The current labels are heavily based on static and system-level activity analysis. 
The problem is that malware family labels do not represent the capabilities of malware samples. The black-box (unexplainable) nature of the labeling methods also makes it impossible to verify assigned family labels, causing 
the evaluation of newer detection methods to depend on unreliable ground truth~\cite{li2010groundtruthproblems}. 
Moreover, network traffic is rarely used to determine family labels because of noisy ground truth and non-stationary data distribution~\cite{anderson2017machine}.  As a result, malware samples that exhibit identical network behavior but have different code attributes end up in different families, see e.g. Perdisci et al.~\cite{Perdiscibehav}. 

In this chapter, we address the limited interpretability of malware family labels by proposing white-box\footnote{In white-box ML, all steps are explainable -- the input, output and how the output was generated. In contrast, only the input and output are known in black-box ML, e.g. Neural Networks.}
behavioral profiles for malware samples. Existing research suggests that network traffic shows malware's core behavior by capturing direct interactions with the attacker or C\&C server~\cite{corebehavior}. Network traffic analysis can also be performed remotely, which presents a lower overhead than many popular system-activity solutions. Therefore, we place emphasis in building network behavioral profiles. 
To this end, we propose MalPaCA (Malware Packet Sequence Clustering and Analysis) for automated capability assessment of malware samples. The goal of \textit{Capability Assessment} is to discover the behaviors a malware sample can exhibit. We investigate the usage of unsupervised machine learning for intelligent capability assessment to tackle the ever-increasing volume of newly discovered malware.

Until now, malware capability assessment has primarily been a manual effort~\cite{12, comparativeMEA, sharma2019malwarecapability}, resulting in behavioral profiles that are quickly outdated. Although machine learning based behavioral analysis approaches exist, they construct a single model that describes either the whole network or each protocol usage individually~\cite{rieck2011automatic}. However, the network traffic originating from even a single host can be so complex that these models fail to correctly represent malicious behaviors~\cite{stratosphere}. This is why MalPaCA splits the network traffic between hosts into \textit{uni-directional connections} and considers them as discrete behaviors (or \textit{capabilities}). 

MalPaCA clusters similar connections based on their temporal similarity, where each cluster represents a unique capability. A malware sample is then represented by its \textit{Behavioral Profile} --- a list of cluster membership of its connections. We represent malware's behavioral profiles in a Directed Acyclic Graph that shows different samples' overlapping behaviors. The graph also shows malware samples from different families behaving identically, showing potentially incorrect family labels. 
MalPaCA is novel as it adopts sequential features that keep the temporal nature of the traffic intact. It uses a combination of Dynamic Time Warping and N-grams to measure the distance between network connections.
MalPaCA utilizes only 20 packets to identify the network behavior shown by any given connection. It also utilizes only the packet header features that are available even when traffic is encrypted. 

The last step of MalPaCA's pipeline is assigning capability labels to clusters. Each discovered cluster is visualized using \textit{temporal heatmaps} to determine which capability it captures. The temporal heatmaps provide a goal- and data-driven  approach to investigate the performance of MalPaCA's clustering, by clearly showing the network connections that are grouped together. This eliminates the need to manually investigate thousands of network traces.
Security analysts can also fine-tune MalPaCA's parameters by visualizing the temporal heatmaps.
The key advantage of this methodology is its white-box and explainable  nature: it provides a visual representation to investigate MalPaCA's rationale for finding behavioral similarity. In doing so, we address the interpretability problem of typical black-box analysis methods, which is an important stepping stone towards better detection methods.

We evaluate MalPaCA's performance on 1.1k malware samples  (resulting in 3.6M packets) coming from 15 families collected in the wild. We also compare the effectiveness of sequence clustering by comparing with an existing method based on frequently-used statistical (aggregate) features~\cite{botfinder}. 

\runinhead{Results. } 
The results are very promising:
(i) MalPaCA's capability assessment works on low quality datasets with as low as 20 packets in each trace, though additional traces result in more thorough profiles;
(ii) It successfully discovers several attacking capabilities, such as port scans and reuse of C\&C servers;
(iii) MalPaCA demonstrates the effectiveness of sequence clustering by producing an error rate of 8.3\% compared to 57.5\% obtained from statistical features; and
(iv) MalPaCA uncovers multiple discrepancies between behavioral clusters and family labels. We believe this happens either because the labels are incorrect or because the overlapping families share significant behavior. 

\runinhead{Contributions. }
We summarize our contributions as follows:
\begin{enumerate}
        \item We show that short sequences of packet header features are capable of characterizing network behavior;
        \item We build \textit{MalPaCA}\footnote{https://github.com/azqa/malpaca-pub} --- a tool to automatically build network behavioral profiles of 
        malware samples collected in the wild; 
        \item We introduce \textit{temporal heatmaps} --- a data-driven and visualization-based cluster evaluation method that requires no ground truth;
        \item We show the behavioral relationships between malwares using a Directed Acyclic Graph, which also uncovers discrepancies between behavioral clusters and traditional family labels;
        \item We demonstrate the effectiveness of sequence clustering, which shows less errors than an existing solution based on statistical features.
\end{enumerate}

\section{The problem with AV labels} \label{sec:av-label-analysis}
This section presents an analysis of our experimental dataset to emphasize the problem of inconsistent AV labels, and motivate the need for explainable behavioral profiles. We compare the \textit{agreement rate} of two popular malware labeling practices, i.e. YARA rules\footnote{https://virustotal.github.io/yara/} and VirusTotal\footnote{https://www.virustotal.com/} labels. %
The malware collection process is given in Section \ref{sec:data}. Table \ref{tab:datasetdistro} shows the number of binaries in each malware family. 

The malware binaries in the dataset are labeled using YARA rules. Each malware binary also has a Virus Total (VT) scan report. On average, there are 61 AV vendors for each malware sample, out of which 25.8\% vendors per malware sample return a \texttt{null} detection, i.e. unable to detect it as malicious. The rest assign various labels to each malware binary.

\begin{figure} [ht]
    \centering
    \includegraphics[scale=.50]{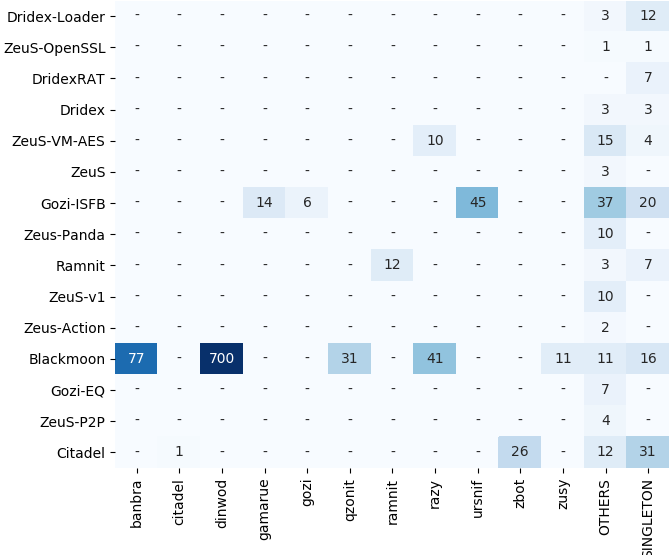}%
    \caption{Disagreements between AV vendors. Rows: YARA labels, Columns: AVClass labels, Counts: \# malware binaries.}
    \label{fig:disagreements}%
\end{figure}
Since each AV vendor has its own vocabulary, a trivial filtering attempt on a VT report cannot identify the true underlying family label. Sebastian et al.~\cite{sebastian2016avclass} have developed an open source tool, called AVClass, that takes VT reports as input and returns the most likely family label. If, after all the filtering steps, AVClass is unable to identify the family name, it declares the malware as a ``SINGLETON". We use AVClass to reduce a VT report into its representative VT family label. In the experimental dataset, AVClass returns ``SINGLETON" for 101/1196 (8.4\%) VT reports, while assigning 42 unique family labels to the rest 1095 malware binaries.

Figure \ref{fig:disagreements} shows the label agreement rate between the YARA and VT labels. The y-axis shows the YARA labels. The x-axis shows the VT labels as aggregated by AVClass. For brevity, ``OTHERS" category contains all samples for which $counts < 10$. 
Only 3 family names co-exist in both YARA and VT labels, i.e. \texttt{Citadel}, \texttt{Gozi}, and \texttt{Ramnit}. Also, although \texttt{Ramnit} is detected under the same name by both YARA and VT, 10 malware samples are still labeled differently. In fact, YARA family labels are assigned 4.2 distinct VT labels on average, while VT labels are assigned 1.5 distinct YARA labels on average. One example demonstrating this is: \texttt{YARA}: \texttt{Zeus-VM-AES} (29 samples) are predicted as \texttt{VT}: \texttt{razy} (10 samples), \texttt{gamarue} (6 samples), \texttt{cerber} (3 samples), \texttt{upatre} (3 samples), \texttt{farfli} (1 samples), \texttt{locky} (1 samples), \texttt{hpcerber} (1 samples) and \texttt{SINGLETON} (4 samples). 
This makes it very hard to understand the collected malware. One fair conclusion is that some VT labels can be considered as sub-families of the popular YARA malware family. For example, \texttt{Dinwod} and \texttt{Banbra} seem to be sub-families of \texttt{Blackmoon}, but the names alone do not explain which attributes set them apart from each other.

%% file: Sec-Litrev.tex
\section{\textbf{Related work}}\label{sec:litrev}
The field of malware analysis has existed since the first malware was discovered over thirty years ago. Since then, multiple machine learning based approaches have been proposed to automate malware detection and analysis. In this section, we present a brief survey of the major research 
challenges targeted by prior work. In doing so, we highlight how our work fills the gaps across various research themes. 

\subsection{Challenges in malware labeling}

Existing research has repeatedly shown that malware family labels are noisy and inconsistent. Popular tools, such as VirusTotal, run multiple AV scanners and return an array of labels predicted by each scanner, without any indication as to which is correct. There is also an absence of a common vocabulary that all security companies can follow to label malware samples. Maggi et al.~\cite{maggi2011finding} propose a method to find inconsistencies in malware family labels generated by Anti Virus (AV) scanners. Mohaisen et al.~\cite{mohaisen2013towards} are the first to measure the accuracy, consistency and completeness of AV scanners. Their results show that AV vendors produce inconsistent labels 50\% of the time, on average. These findings resulted in research that found ways to deal with the inconsistencies in the family labels. Kantchelian et al.~\cite{kantchelian2015better} proposed an algorithm based on Expectation Maximization and Bayesian models that assigns weights to each vendor's trustworthiness. Sebasti{\'a}n et al.~\cite{sebastian2016avclass} developed a useful open source tool, called AVClass, that determines the likely family name after performing heavy filtering on all the predicted labels. However, these methods do not address the key underlying issue---malware family labels are black-box with limited interpretability. 

\textit{Behavioral profiles} complement family names in that they also describe the behavior of a sample. 
\textit{Capability assessment} is done to characterize a malware family, 
which has primarily been a manual effort resulting in behavioral profiles that are quickly outdated. Also, most of the prior works in capability assessment utilize information extracted from the static analysis of malware executables: Black et al.~\cite{12} bridge the semantic gap between low-level API calls and high-level behaviors in order to build a taxonomy of banking malware. They extract API calls by statically analyzing a banking malware dataset, and map them to high-level behaviors manually with the help of domain experts. Sharma et al.~\cite{sharma2019malwarecapability} recently proposed a method to automatically build behavioral profiles. They select a few high-level capabilities possessed by malware by investigating the literature, and map them to low-level behaviors extracted from the static analysis of 56 malware samples. 
\textit{In contrast, we propose MalPaCA that automatically builds dynamic (network) behavioral profiles. 
}

\subsection{Research Objectives: Detection versus Analysis} 
Existing research on malware comes in two strains: detection-based and analysis-based.
Malware detection and signature generation dominates existing literature, with the end-goal of optimizing metrics~\cite{17, 22, bayer2009scalable, Perdiscibehav, botfinder, garcia2009anomaly, lin2018tabor, wei_wang_malware_2017, rafique2013firma, malwarevariants, acar2018peek, conti2015can, aiolli2019mind}; while only a few of these works also help the readers understand and analyze the obtained results~\cite{10, stratosphere}.
Recently, however, several malware analysis approaches have been proposed that aim to improve malware understandability rather than optimizing detection rates. These methods provide essential insights that can improve malware detection methods. Black et al.~\cite{12} perform an in-depth analysis of the key behaviors of banking malware families and how they have evolved over time.  
Moubarak et al.~\cite{comparativeMEA} discuss malware evolution and the structural relationship between several potentially state-sponsored malware. 
In~\cite{dendroid}, the authors
cluster Android malware samples, and build a dendrogram of the malware families showing overlapping code snippets.  
Sharma et al.~\cite{sharma2019malwarecapability} build behavioral profiles of malware samples using static analysis. \textit{In this chapter, we follow a similar approach and build an analysis tool, MalPaCA. MalPaCA uses unsupervised clustering to group network connections that behave similarly, and uses them to construct malware's behavioral profiles.}

Although clustering is an unsupervised technique, existing literature has often used some notion of ground truth (family labels) to evaluate the cluster quality. 
Bayer et al.~\cite{bayer2009scalable} evaluate their malware clustering approach using labels obtained by the majority voting of 6 AV vendors. 
Perdisci et al.~\cite{Perdiscibehav} evaluate their malware clustering approach by introducing a notion of AV graphs that depict the agreement between AV vendors as a measure of cluster cohesion and separation. 
In~\cite{malwarevariants}, the authors report the precision and recall of higher than 0.95 of their malware clustering approach. They use the majority voted family labels from 25 AV vendors as their ground truth. Li et al.~\cite{li2010groundtruthproblems} have examined the challenges of evaluating malware clustering, and have advised caution when deciphering highly accurate clustering results as they can be impacted by spatial bias: performing majority voting on AV-provided labels is hazardous, since if most of the AV vendors are in agreement, it typically indicates that the families are already easy to detect. \textit{In this chapter, we propose a data-driven and visualization-based method to evaluate 
clusters, without using family labels. Instead of optimizing clustering accuracy, our emphasis is on explainability of the results.}

\subsection{Challenges in malware behavior modeling}
Modeling software behavior is  
a challenging task, but modeling malware's behavior is even more challenging since malware authors specifically try to evade detection~\cite{chakkaravarthy2019survey}.
Static analysis of malware binaries and disassembled code has been a popular malware analysis approach in the literature~\cite{feng2014apposcopy, 12, 22, malwarevariants, azab2014mining}. Increasingly more malware uses obfuscation techniques to evade analysis, causing difficulties for statically analyzing malware. 
The obfuscation attempts gave rise to dynamic analysis of 
malware that executes a malware sample in a sandbox and collects execution traces from it. Dynamic analysis is generally divided in two strains: System activity and Network traffic analysis.
Network traffic analysis collects traces of malware samples remotely using existing network monitoring infrastructures~\cite{Perdiscibehav}, making it much easier to apply.
However, the behavioral analysis and signature generation literature is heavily focused on system activity analysis, e.g. see~\cite{bayer2009scalable,9, scannerthesis, sharma2019malwarecapability}. Research suggests that network traffic shows the core behavior of malware~\cite{corebehavior}. Although sometimes encrypted, network traffic contains the direct interaction with the attacker.
In this section, we discuss three major challenges of modeling malware behavior via traffic analyses.

\runinhead{Feature selection.}
Network traffic analysis is generally applied when 
designing Network Intrusion Detection Systems (NIDS), which either detect anomalous  traffic~\cite{garcia2009anomaly} or generate signatures for malware families~\cite{18,toomuchmalware,tian2009automated}.
Deep Packet Inspection (DPI) is one commonly used approach in NIDS to extract information from packet payloads. 
For example, Rafique et al.~\cite{rafique2013firma} use DPI for automatic signature generation of malware families. 
Although effective, downsides to DPI-based approaches are that they are privacy-intrusive, operationally expensive, and do not work out-of-the-box for encrypted traffic.
There are also 
approaches that 
detect specific attacks.
For example, HTTP-based malware can be detected using specific features from the Application header 
~\cite{Perdiscibehav}. Similar approaches exist for DNS-based malware  
~\cite{21,dnsbased}, and HTTPs-based malware 
~\cite{https1}. In the absence of the HTTP, DNS and TLS headers, these approaches seize to work.

Several works use \textit{coarse} or high-level features 
that are protocol-agnostic and work out-of-the-box even with encrypted traffic.
For example, Conti et al.~\cite{conti2015can} use sequences of packet sizes to characterize the network behaviors generated by Android applications.
Aiolli et al.~\cite{aiolli2019mind} use various statistical features computed over packet sizes to detect bitcoin wallet application functionality.
Acar et al.~\cite{acar2018peek} use network traffic direction and packet lengths to identify commands issued to smart home IoT devices. 
These works aim to characterize benign network behaviors.
In the malware domain, 
Tegeler et al.~\cite{botfinder} use average packet size, average packet inter arrival time, average connection duration and the FFT of C\&C communication to detect bot infected  hosts. 
Garcia~\cite{stratosphere} builds a behavioral Intrusion Detection System by using the size, duration and periodicity of Netflows. 
\textit{In this chapter, we also use high-level features from packet headers to characterize malware's network behavior. To the best of our knowledge, network traffic analysis has not been used in capability assessment or for generating behavioral profiles of malware samples.}

\runinhead{Feature representation. }
Machine learning methods take a feature vector as input, 
which can represent anything ranging from a single behavior to a complete malware sample. 
Multiple observations for a single feature are aggregated into \textit{statistical features}, e.g. mean packet size of a netflow. 
Existing literature is filled with approaches that use such statistical features, e.g., see~\cite{stratosphere, 17, botfinder, azab2016machine}.
Although they are computationally efficient, 
they lose local behavioral details, which can be a problem when the goal is to characterize that behavior.

Another approach that is gaining momentum is the use of \textit{sequential features}. Numeric sequential features are typically used in two ways: \textit{Discretized} and \textit{Raw sequences}. A raw sequence (or a continuous sequence) is composed of the original observations; while a discretized sequence encodes the observations into a finite set of bins.
 Discretizing sequences 
 is typically faster and makes measuring distances easier. Pellegrino et al.~\cite{10} learn state machines from discretized netflow data in order to detect bot-infected traffic, while Hammerschmidt et al.~\cite{hammerschmidt2016behavioral} use it to cluster host behavior over time.
Lin et al.~\cite{lin2018tabor} detect anomalies in industrial water treatment plant by using discretized sequences from sensor readings.  
In practice, malware-related data is often scarce and noisy. In this case, discretization can lose important information. 

Raw sequences are rarely used for modeling network traffic because it is non-stationary and contains noise (e.g. empty acknowledgment packets or retransmissions), and delays (due to varying network latency)~\cite{anderson2017machine}. 
Ntlangu et al.~\cite{ntlangu_modelling_2017} provide a brief overview of time-series approaches to model network traffic.
As noted in~\cite{ntlangu_modelling_2017}, due to the nature of network traffic and their distributions, (auto-)regressive models struggle to accurately capture them.
Kim et al.~\cite{kim_android_2015} use a multi-variate time-series regression model on host-based resource consumption, such as CPU and memory usage (not network traffic)
to identify Android malware.
Conti et al.~\cite{conti2015can} propose a method to detect the action performed by Android applications using raw sequential features.  
\textit{To the best our knowledge, MalPaCA is the first method that successfully uses short raw sequential features 
to characterize malware network behavior.
}

\runinhead{Distance measure.}
The notion of behavioral similarity requires the means to be able to measure distance between two objects. 
The choice of the distance measure is directly dependent on the data type of the feature set (e.g. numeric or categorical) and the way the features are represented (e.g. statistical or sequential). 
For statistical features, Euclidean distance is most commonly used.
For instance, Chan et al.~\cite{scannerthesis} use Euclidean distance to determine similar Android processes.

Calculating the distance between sequential features is more challenging because they may not always be properly aligned.
For categorical (or discretized) sequences, there exist Bioinformatics inspired solutions using sequence alignment~\cite{vinod2012seqalignment}. They require pre-computed substitution matrices, which currently do not exist for malware. 
There also exist String matching solutions frequently used in the Natural Language Processing domain. 
Baysa et al.~\cite{baysa2013structural} use Levenshtein, or edit distance, to measure the similarity between two malware binary files.
A sequence can also be broken down into sub-sequences, represented as Ngrams, which
have been used to model genomic sequences~\cite{dnangram} and to match files~\cite{fileprintsngram}. They have also been used to classify malware families in~\cite{canfora2015effectiveness}.
Longest Common Subsequence (LCS) with \texttt{k}-gaps can also be used to measure distances between sequences. The gaps account for the occasional noise. Chan et al.~\cite{scannerthesis} use LCS to group similar resource-access-patterns (not network traffic) in Android applications. 

A few distance measures exist for raw or continuous sequences. 
Verwer et al.~\cite{verwer2014pautomac} have used Kullback-Leibler divergence to measure the distance between two sequences while learning probabilistic automata. However, it requires substantial amount of data to measure the similarity with a high confidence, which is not always available for malware. 
Another promising distance measure is Dynamic Time Warping (DTW). 
DTW has been used in fingerprint verification~\cite{fingerprint}, characterizing DDoS attack dynamics~\cite{ddosdynamics}, and measuring similarity in android application behavior~\cite{conti2015can}.
\textit{MalPaCA uses a combination of DTW and Ngrams to measure the distance between network connections.
}

%% file: Sec-Method.tex
\section{\textbf{MalPaCA: Malware Packet Sequence Clustering and Analysis}}\label{sec:method}

The ultimate goal of MalPaCA is to construct a behavioral profile for each malware sample that is more descriptive than its family label. Research shows that malware belonging to the same family exhibits similar behaviors since malware authors often share code and resources~\cite{20}. To this end, MalPaCA automatically identifies the various network behaviors exhibited by malware samples, and groups samples that share common behavior. MalPaCA does not assume any a priori knowledge about the malware's family name or its capabilities, and hence can be used out-of-the-box for other malware datasets. The profiles are built using observed behavior since only the executed functionality is relevant for behavioral profiling. Profiles for individual families can be enriched further by observing additional traffic. We release MalPaCA to the public\footnote{https://github.com/azqa/malpaca-pub}.

\begin{figure*}[t]
    \centering
    \includegraphics[width=\linewidth]{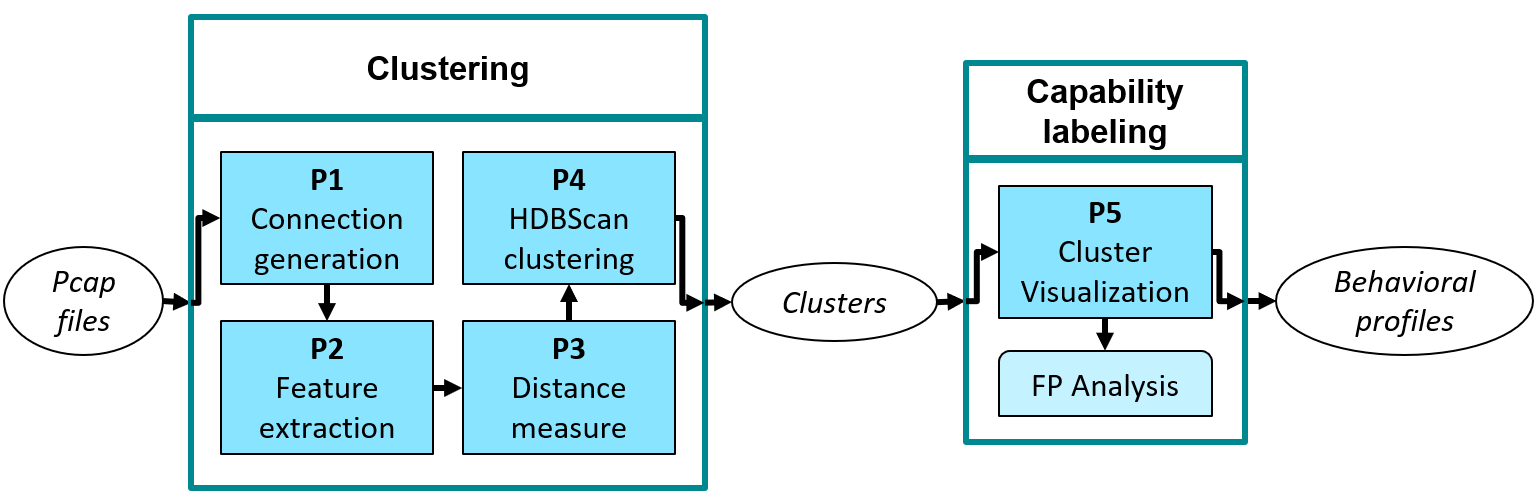}
    \caption{MalPaCA: Connections clustered on behavioral similarity; malware described using connections' cluster membership.}
    \label{fig:workflow}
\end{figure*}

Figure \ref{fig:workflow} illustrates the architecture of MalPaCA with its five phases (P1 to P5).
Network traces (Pcap files) are given as input to the system, which are split into unidirectional packet streams (or \textit{connections}) that are clustered based on temporal similarities.  
Each cluster is assigned a capability label by visualizing temporal heatmaps showing connections' feature values. {Each malware sample (and its associated Pcap file) is then described by a \textit{Cluster Membership String}, forming a descriptive behavioral profile.}

\subsection{\textbf{Connection generation} (P1)}
A \textit{connection} is defined as an uninterrupted uni-directional list of all packets sent from source IP to destination IP address. This means 8.8.8.8 $\rightarrow$ 123.123.123.123 is a different connection than 123.123.123.123 $\rightarrow$ 8.8.8.8. 
We refer to these as \textit{Outgoing} and \textit{Incoming} connections based on their direction with respect to the \texttt{localhost}. Note that we do not use IP address as a feature, except to create connections.

Ideally, a connection captures one complete capability.
 The connection length can vary significantly depending upon the behavior and network delays. Since the network delay is an artifact of the network, not of the malware, it is important to reduce its impact when measuring behavioral similarity. MalPaCA does so by capping 
 the sequence length to a fixed threshold, avoiding artifacts that are due to connection length.

 Existing research suggests that it is possible to identify behavioral differences from a \textit{handshake}\footnote{{\textit{Handshake traffic refers to the introductory few packets of a connection.}}}. 
 Wang et al.~\cite{16} use the first 3 to 12 bytes of packet headers in order to identify the different so-called Protocol Format Messages.
MalPaCA builds upon this idea and utilizes the first few packets of a connection to identify the capability. This is a fixed threshold denoted by the tunable parameter $len$.
It should be large enough to allow the handshake to be modeled, the length of which is often unknown in network traffic analysis.
Larger values of $len$ not only include noise artifacts but also increase the computational resources required to process longer connections.

\subsection{\textbf{Feature-set extraction} (P2)}
The choice of feature-set is crucial for determining the kind of behaviors that are identified by MalPaCA. Two considerations motivate our choice: 
1) MalPaCA should be generalizable to more than one type of malware;
2) The feature set is small and easy to extract. Hence, we do not use features extracted from the packet payload itself as they limit the applicability of the method. We also do not use IP addresses as they are easy to spoof and are considered Personally Identifiable Information\footnote{https://www.enterprisetimes.co.uk/2016/10/20/ecj-rules-ip-address-is-pii/} in countries like the Netherlands. 
We use four sequential features: (i) packet size, (ii) time interval, (iii) source port, (iv) destination port.
All four features are independent of the protocol type, making them available for every connection. Each feature is represented as a sequence of raw observations for subsequent packets. Although these features are simplistic, we demonstrate that their sequential nature captures malware behavior effectively. 

\emph{Packet size ($f_{ps}$)} measures the size of the \textit{IP datagram} of each packet in bytes. 
\emph{Time interval ($f_{in}$)} captures the inter packet arrival time in milliseconds.
We use time interval because malware tends to show a periodic behavior, e.g. bots send periodic heartbeat packets\footnote{https://www.ixiacom.com/company/blog/mirai-botnet-things} to inform the C\&C server about the infected host. MalPaCA is meant to be used on a single network at a time since using inter-arrival time makes connections collected on different latency networks incomparable. 

We use both \emph{source ($f_{sp}$)} and \emph{destination ($f_{dp}$)  port numbers} because the connections are unidirectional. We particularly use source port so the analysts can limit the use of problematic ports in case of outgoing connections.
The usage of certain vulnerable ports can also indicate suspicious activity.
Each connection is represented by four sequences, one per feature, $C = (f_{ps}, f_{in}, f_{sp}, f_{dp})$.

\subsection{\textbf{Distance measure} (P3)} \label{sec:dist}

Three considerations motivate our choice of distance measure: 1) Different distance measures are applicable on numeric and categorical data types; 2) The distance measure should be intuitive to help understand the results; 3) It must produce results that are resilient to delays and noise, which are common characteristics of network traces. The last consideration was added after observing distance measures producing results that were artifacts of network delays.
MalPaCA uses a combination of Dynamic Time Warping (DTW) and N-gram analysis to measure distance between two connections. 

\runinhead{Dynamic Time Warping}. DTW~\cite{dtw} is used to measure distances between numeric sequences (packet size and time interval) due to its robustness to delays and noise. 
It aligns two time-series that may contain distortions (or warps) in the time-axis. It maps local substructures in one sequence to those of the other sequence. 
For two sequences $a = [a_0, a_1, ..., a_n]$ and $b = [b_0, b_1, ..., b_m]$
the DTW distance $d_{dtw}(a,b)$ is:

\begin{equation}\label{eq:dtwformula}
d_{dtw}(a,b) = 
\sum_{i=1}^{n+1}\sum_{j=1}^{m+1} ||a_i-b_j|| + \min \begin{cases} d(a_{i-1}, b_j), \\
d(a_i, b_{j-1}), \\
d(a_{i-1}, b_{j-1})) \end{cases}
\end{equation}

The output is a \textit{similarity} score, which we normalize using:

\begin{equation}\label{eq:norm}
d_{ndtw}(a,b) = \frac{d_{dtw}(a,b) - \min_{x,y}\{d_{dtw}(x,y)\}}{\max_{x,y}\{d_{dtw}(x,y)\} - \min_{x,y}\{d_{dtw}(x,y)\}}
\end{equation}

\runinhead{Ngram analysis.} An \textit{Ngram} is defined as the set of $n$ (called $order$
) consecutive items in a given sequence. 
The  larger the value of $order$, the more sequence structure is captured.
A sequence of port numbers is converted into a set of ngrams, called its \textit{Ngram profile} using a sliding window of length $order$. An example for $order=2$ is shown in Table \ref{tab:ngramexample}, where $A,B,C,D$ are hypothetical port numbers. Let $G$ be the set of all unique Ngrams occurring in the dataset. 
For each packet sequence $a$, a vector $a_{g} = [f(g_1,a), f(g_2,a), \ldots , f(g_{|G|},a)]$ is generated, containing the occurrence frequencies $f(g_i,a)$ in $a$ of each Ngram $g_i \in G$.

\begin{table}[ht]
\centering
\renewcommand{\arraystretch}{1.3}
\caption{Example -- Distance measurement using Ngram analysis.}
\label{tab:ngramexample}
\scalebox{0.85}{
\begin{tabular}{c|c|c|c}
\textbf{Input} & \textbf{Ngram profiles} & \textbf{G={[}AB,BC,CB,DA,CA{]}} & \textbf{Cosine distance} \\ \hline
$ABCBC$ & $AB,BC,CB,BC$ & {[}1,2,1,0,0{]} & \multirow{2}{*}{0.3876} \\ \cline{1-3}
$DABCA$ & $DA,AB,BC,CA$ & {[}1,1,0,1,1{]} & 
\end{tabular}}
\end{table}

We measure the distance between two Ngram profiles 
using Cosine distance. Other distance measures exist for Ngrams, 
but Cosine has shown promise in measuring similarity between categorical sequences 
~\cite{zahrotun2016jaccardvscosine}. 
It is determined by the angle between two non-zero vectors. The similarity value lies between 0 and 1, where 1 means that the two vectors are the same (parallel to each other) and 0 means they are completely different (orthogonal to each other). For two sequences in their vector representations $a = [v_1,\ldots,v_{|G|}]$ and $b = [v_1',\ldots,v_{|G|}']$, the Cosine 
distance $d_{cos}(a,b)$ is:

\begin{equation}\label{eq:cosineformula}
d_{cos}(a,b) = 1 - 
\frac{\sum_{i=1}^{|G|}a_i \times b_i}{\sqrt{\sum_{i=1}^{|G|}a_i^2} \times \sqrt{\sum_{i=1}^{|G|}b_i^2}} 
\end{equation}

Finally, the DTW and cosine distances are combined to calculate the final distance between two connections:

\begin{equation}
\label{eq:distconn}
d_{conn}(A,B) = \frac{
d_{ndtw}(a_{ps},b_{ps}) + 
d_{ndtw}(a_{in},b_{in}) + d_{cos}(a_{sp},b_{sp}) + d_{cos}(a_{dp},b_{dp})} {4}
\end{equation}\\
where $A = (a_{ps},a_{in},a_{sp},a_{dp})$ and $B = (b_{ps},b_{in},b_{sp},b_{dp})$ are connections and their features: packet sizes $\{a|b\}_{ps}$, intervals $\{a|b\}_{in}$, source port Ngram profiles $\{a|b\}_{ps}$, and destination port Ngram profiles $\{a|b\}_{dp}$.

\subsection{\textbf{HDBScan Clustering} (P4)}
 
 A key strength of MalPaCA is the clustering algorithm it uses.
 There exists a familial structure among malware behaviors~\cite{tian2009automated,dendroid}. 
 Therefore, it makes sense to use hierarchical clustering to model the relationships between them. 
 We have used Hierarchical Density-Based Spatial Clustering of Applications with Noise (HDBScan)~\cite{hdbscan} for this purpose. The key strengths of HDBScan are twofold: it automatically determines the optimal number of clusters, and it generates high-quality clusters that remain stable over time. 
 It also has minimal tunable parameters, which allow configurations to be generalizable. 
 
 HDBScan requires a pairwise distance matrix as input. It does not force data points to become part of clusters---all data points whose membership to a cluster cannot be determined are considered to be \textit{noise}. 
 In our context, \textit{noise} refers to behaviors that are either too different from all the others or cannot be clearly assigned to one cluster.
 An ideal dataset with clear cluster boundaries will have no noise.
 Hence, in the presence of a less ideal dataset, noise is discarded to extract high-quality clusters. Keep in mind that discarding excessive connections as noise can also be counterproductive. We discuss this limitation in Section \ref{sec:lim}.

\subsection{\textbf{Cluster visualization} (P5)} \label{Sec:clusteranalysis}
Formalizing cluster quality without ground truth is a fundamental challenge in clustering. Although some metrics exist that capture cluster quality (i.e. Silhouette index~\cite{silhouettes} and DB Index~\cite{dbindex}), they require a notion of distance from a cluster centroid, which is difficult to obtain for sequences.
In MalPaCA, each connection is represented by four sequences and collapsing these into a single cluster quality measure loses important local behavior. Instead, 
we define the following properties to be indicative of good clustering:
(1) Cluster homogeneity is high---a cluster contains only similar connections.
(2) Cluster separation is high---each cluster captures a unique capability.
(3) Clusters are small and specific so they only capture the core capability.
The first two properties ensure that we obtain meaningful capability-based clusters, the third ensures that only the core capabilities are captured.

We use \textit{temporal heatmaps} for a white-box cluster analysis. We graphically show the connection features and rely on human visualization skills to determine cluster quality. Analysts can inspect heatmaps to determine which behavior is captured in a cluster. This gives them control over the clustering results. 
We leave the automation of this process as future work.

\begin{figure*} [t]
    \centering
    \subfloat[Packet sizes]{{\includegraphics[width=0.25\linewidth]{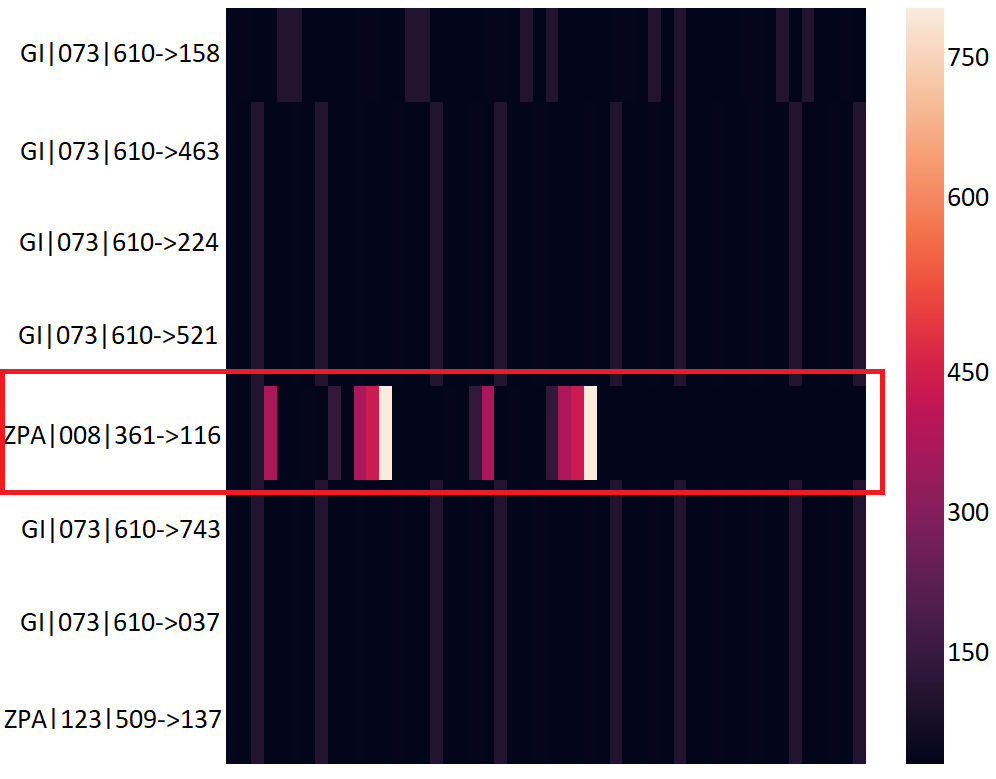}}}%
    \subfloat[Interval]{{\includegraphics[width=0.25\linewidth]{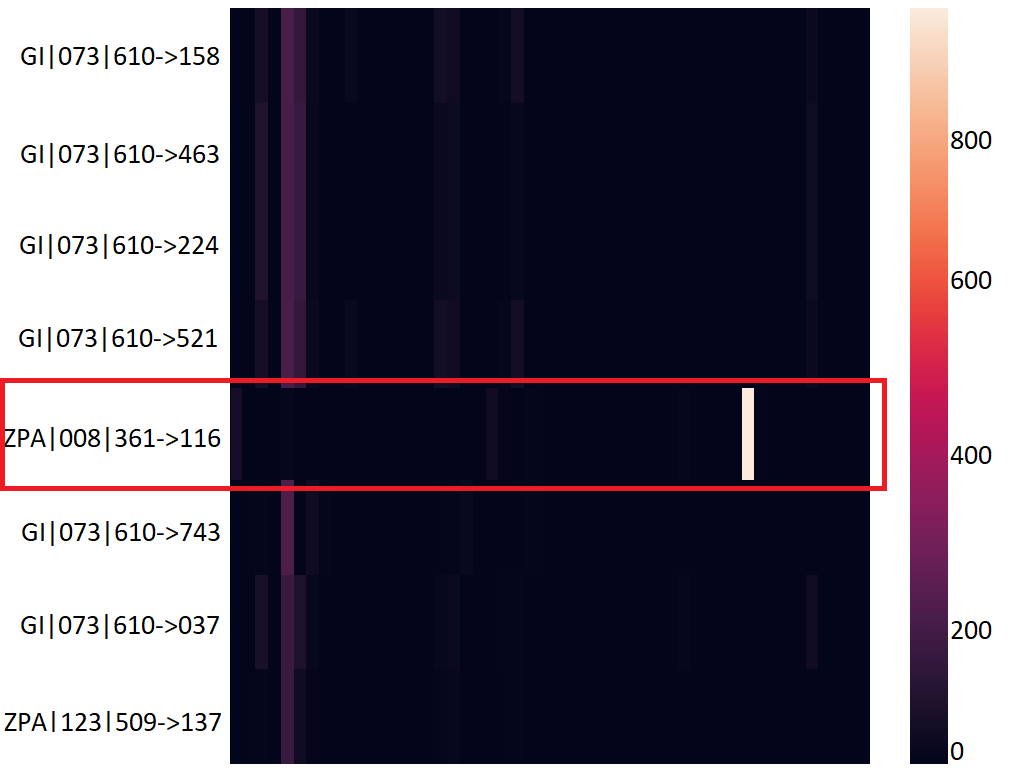}}}%
    \subfloat[Source Port]{{\includegraphics[width=0.25\linewidth]{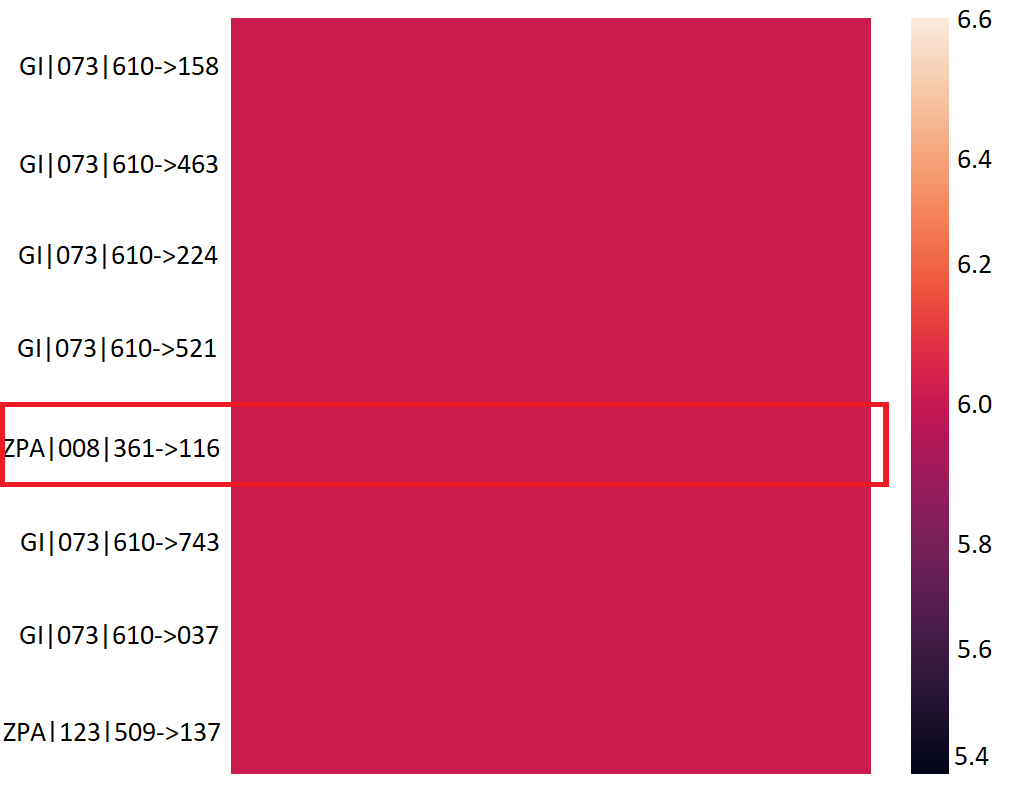}}}%
    \subfloat[Destination Port]{{\includegraphics[width=0.25\linewidth]{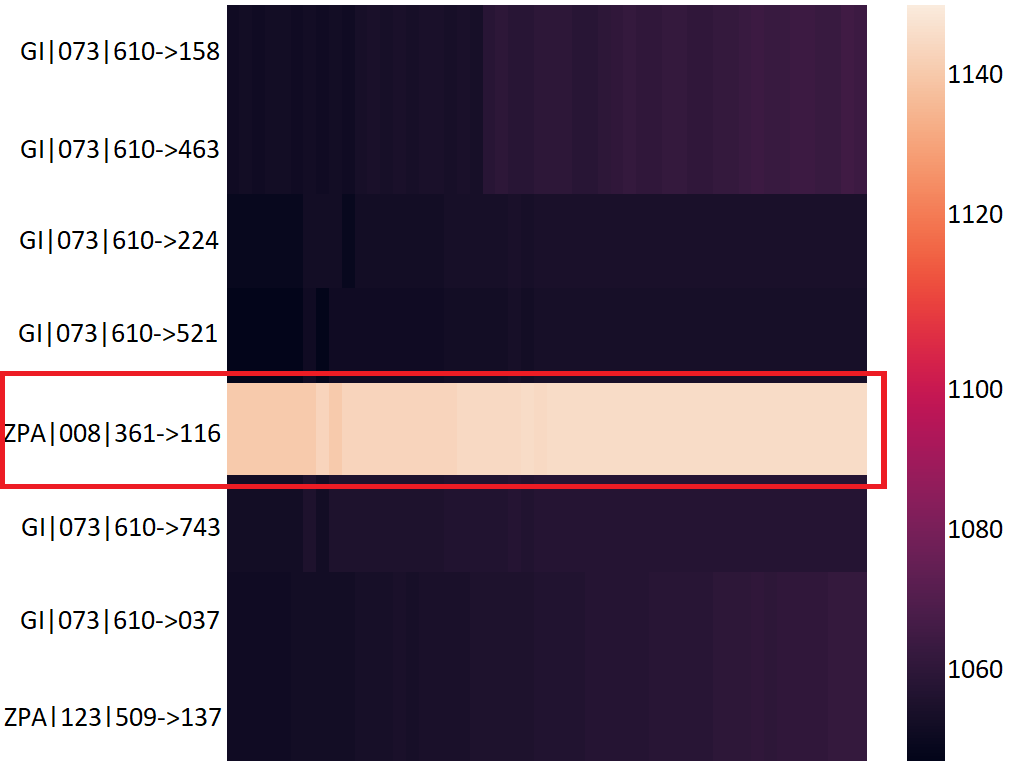}}}%
    \caption{A clustering error: one connection does not belong in the cluster it is assigned}%
    \label{fig:fp}%
\end{figure*}

Four temporal heatmaps are associated to each cluster, one corresponding to each feature. 
Each row in a heatmap shows the corresponding feature sequence of the first $len$ packets in a connection. 
Figure \ref{fig:fp} shows example temporal heatmaps.
The figure highlights one dissimilar connection among the eight in the cluster, clearly highlighted in red.

\runinhead{Clustering Error Analysis.}
Visualizing the cluster content helps to identify which connections belong in a cluster.
A \textit{Clustering Error} (CE) is defined as a connection that is placed in cluster \texttt{X} despite half of its features being different from the remaining connections in the cluster. 
Since each feature holds equal weight, we only consider a connection as CE if more than two features \textit{differ}. 
We consider two features \textit{different} if more than 50\% of their sequences differ so significantly that a different color appears on the temporal heatmap. This is where human visualization skills play a key role in determining feature similarity. Figure \ref{fig:fp} shows a cluster containing one CE, highlighted in red. It shows that three out of four feature values of this connection are different from other connections in the same cluster. The clustering error rate is calculated as $\frac{CEs}{Cluster size}$, i.e., $\frac{1}{8}$. We measure the error rate of each cluster similarly, and calculate the \textit{average percentage of errors per cluster} as a notion of clustering quality. 

In practice, we first establish the common majority by finding two or more connections that are most similar to one another, i.e., the ones that have the least mutual distance. The pair-wise distance matrix computed during clustering is used as a lookup table for finding such connections. 
Figure \ref{fig:fp} shows a simple case where the \textit{rightful owners of a cluster} are easily visible since 7 out of 8 connections are very similar. 
The rest of the connections are compared with the rightful owners and are either considered as true positives or clustering errors, depending on how many feature sequences differ. 

\section{\textbf{Experimental Setup}} \label{sec:exp}
In this section, we describe the dataset used for the experiments and the configuration details of MalPaCA's parameters. 
\subsection{\textbf{ Experimental Dataset}} \label{sec:data}
MalPaCA was evaluated on  financial  malware samples collected in the wild. We worked in collaboration with a security company that specializes in malware analysis and threat intelligence. They collected the dataset independently. The dataset contained 1196 malware samples that were collected over one year. Each malware sample was executed in a sandboxed environment containing several virtual machines. The resulting network traffic was stored in a Pcap file. 
Some samples showed sandbox evasion. They were re-executed in a VM with different settings. This resulted in a total of 1196 Pcap files. Uni-directional connections were extracted, 
resulting a total of 8997 connections containing 3.6M packets. 

The dataset contains 15 famous financial  malware families. They were labeled by the security company using their proprietary YARA rules. 
Additionally, each sample was submitted to VirusTotal (VT), which hosts 68 AV vendors. For each sample, VT returns a report containing detection results from each vendor. 
Table \ref{tab:datasetdistro} summarizes the dataset. %

\begin{table}[t]
\caption{Experimental Dataset: Malware binaries and their associated YARA family labels.}
\label{tab:datasetdistro}
\centering
\begin{tabular}{lr@{\hskip 2pt}l}
\toprule
\textbf{Family name (YARA)} & \multicolumn{2}{c}{{\makecell{\bf \# Malware binaries} }} \\ \midrule
Blackmoon (B) & 887 & (74.10\%)  \\ 
Gozi ISFB (GI) & 122 & (10.19\%)  \\ 
Citadel (C) & 70 & (5.85\%)  \\
Zeus VM AES (ZVA) & 29 & (2.42\%)   \\
Ramnit (R) & 22 & (1.83\%)   \\
Dridex Loader (DL) & 15 & (1.25\%)   \\ 
Zeus v1 (Zv1) & 10 & (0.83\%)  \\ 
Zeus Panda (ZPa) & 10 & (0.83\%) \\ 
Gozi EQ (GE) & 7 & (0.58\%) \\
Dridex RAT Fake Pin & 7 & (0.58\%) \\
Dridex (D) & 6 & (0.50\%) \\
Zeus P2P (ZP) & 4 & (0.33\%) \\
Zeus (Z) & 3 & (0.25\%) \\
Zeus OpenSSL & 2 & (0.17\%) \\
 Zeus Action & 2 & (0.16\%) \\
\midrule \midrule
\textbf{Total} & \textbf{1,196} & (100\%) \\ \bottomrule
\end{tabular}

\end{table}

\subsection{\textbf{MalPaCA Parameters}} \label{sec:params}
MalPaCA has four parameters, i.e. $order$ of the Ngrams used for port numbers, $len$ of packet sequences for features, and the two parameters of HDBScan clustering algorithm: $Minimum\_Cluster\_Size$ and $K\_nearest\_neighbors$.
In our experiments, we have used trigrams ($order=3$) for port numbers, because they form a good trade-off between performance and data sparsity 
~\cite{lengthngrams}. In the experimental dataset, the length of connections is highly skewed towards shorter sequences, with a mean of 20 packets. We use this mean as $len$\footnote{\textit{$len$ can be adjusted based on the required behavioral specificity.}}. Out of 8997 connections in the dataset, 733 connections are longer than $len$. The HDBScan algorithm uses $Minimum\_Cluster\_Size = 7$ and $K\_nearest\_neighbors = 7$. These parameters were selected by tuning MalPaCA on a configuration dataset (5\% of the usable data). The experiments were run on a machine with Intel Xeon E3-12xx v2 processor, 8 cores and 64GB RAM.

The specificity of the identified behaviors is highly dependent on the length of sequences, i.e. $len$. Based on preliminary experiments with $len=\{5, 10, 20, 50\}$, we found that $len=20$ provided the optimal trade-off between behavior characterization and the amount of connections that were discarded. For smaller values, the connections were too generic. For larger values, connections with slight behavioral differences were considered very different. For example, at $len=50$ several clusters capture slightly different variations of port scans, while at $len=20$ those variations merge to form a few strong clusters.

%% file: Sec-Results.tex
\begin{table}[t]
\caption{For each cluster, (i) \# connections, (ii) \# malware families, (iii) Capability label, and (iv) Traffic direction.}
\label{tab:rarecommon}
\centering
\begin{tabular}{lcccc}
\toprule 
\textbf{Cluster} & \textbf{\# Conns} & \textbf{\# Families} & \textbf{Behavior} & \textbf{Direction} \\ \hline 
c1 & 39 & \textit{\textbf{9 (Common)}} & SSDP traffic & Out \\ 
c2 & 90 & \textit{\textbf{9 (Common)}} & Broadcast traffic & Out \\
c3 & 9 & 4 & LLMNR traffic & Out \\ 
c4 & 49 & 5 & Systematic port scan & In \\ 
c5 & 56 & 5 & Randomized port scan & Out \\ 
c6 & 25 & \textit{\textbf{1 (Rare)}} & Connection spam & In \\ 
c7 & 23 & \textit{\textbf{1 (Rare)}} & Connection spam & Out \\ 
c8 & 16 & \textit{\textbf{1 (Rare)}} & Malicious subnet & Out \\ 
c9 & 11 & \textit{\textbf{1 (Rare)}} & Connection spam & Out \\ 
c10 & 9 & 2 & HTTPs traffic & Out \\ 
c11 & 8 & 2 & C\&C Reuse & In \\ 
c12 & 18 & 4 & HTTPs traffic & In \\ 
c13 & 25 & 5 & Misc. & In \\ 
c14 & 10 & 3 & Misc. & In \\ 
c15 & 20 & 3 & Misc. & In \\ 
c16 & 12 & 3 & Misc. & Out \\ 
c17 & 19 & 3 & Misc. & Out \\ 
c18 & 10 & 4 & Misc. & Out \\ \bottomrule
\end{tabular}
\end{table}

\section{Malware capability assessment} \label{sec:results}

MalPaCA produces 18 clusters from the dataset. There are, on average, 25 connections in each cluster. The algorithm discards 284 connections as noise. The remaining 449 connections originate from 216 Pcap files.
Each cluster captures a unique behavior, listed in Table \ref{tab:rarecommon} along with the malware families that show that behavior. We describe a few of the interesting behaviors obtained by MalPaCA. We also discuss how host-based blacklisting~\cite{ghafir2015blacklist, botfinder}, which is a very common practice in security companies, will fail to detect these behaviors.

\begin{enumerate}
\item\textbf{Connection Direction Identification.}
MalPaCA successfully identifies the direction of traffic flow even though no such feature is used. The clusters and their traffic direction are listed in Table \ref{tab:rarecommon}.  Interestingly, we continue to see this pattern even when port-related features are removed from the clustering. Hence, the sequence of packet sizes and their inter-arrival time are collectively indicative of the flow direction. This important trait identifies whether the suspicious behavior is originating from inside the network or from outside it. 

\item\textbf{Device Probing.} 
Some clusters capture connections that connect to the same host. For example, one cluster contains all connections broadcasting to 239.255.255.250, which is used by the SSDP protocol to find Plug and Play devices. Another cluster captures all connections broadcasting to 224.0.0.252, which is used by the Link-Local Multicast Name Resolution (LLMNR) protocol to find local network computers. These clusters could easily have been obtained by using IP-based blacklist, but they would not have clustered behaviorally similar hosts with different IP addresses. 

\item\textbf{Split-personality C\&C Servers.} In several instances, an infected host was observed responding differently to the same request, so much so that the resulting connections ended up in different clusters. For example, two connections of \texttt{Gozi-ISFB} contact 
46.38.238.XX, which has been reported as a malicious server located in Germany. The outgoing connections are identical as they both request for the same resource. However, the responses received are very different---the first response contains a small packet followed by a series of 1200-byte packets, while the second one contains a periodic list of small and large packets in the range of 600 to 1800 bytes. This insight portrays a better picture of the behavior of said C\&C server. In contrast, a blacklist would have grouped these connections since they belong to the same host.
\item\textbf{Port Scan Detection.} 
Some clusters capture a \textit{Port Scan}\footnote{https://whatismyipaddress.com/port-scan}, which is a method for determining open ports on a device in a network. Port scans are usually a part of the reconnaissance phase in the attack kill chain~\cite{yadav2015technical}. Utilizing sequences of port numbers enables us to detect any suspicious temporal behavior before an attack happens. The clusters identify two types of port scans: (i) \textit{Systematic port scan} where ports are swept incrementally, which is seen as a gradient in the corresponding temporal heatmap; and (ii) \textit{Randomized port scan} where ports are contacted randomly, which shows up in the heatmap as a checkered pattern. See Figure \ref{fig:portscan}. Port scans carried out by different connections are clustered together if they contact the same range of port numbers, which increases their mutual similarity. This result is in direct contrast with Mohaisen et al.~\cite{22} who conclude that port numbers are the least useful features in distinguishing malware families.

\begin{figure} [t]
    \centering
    \subfloat[Systematic port scan.]{{\includegraphics[width=0.5\linewidth]{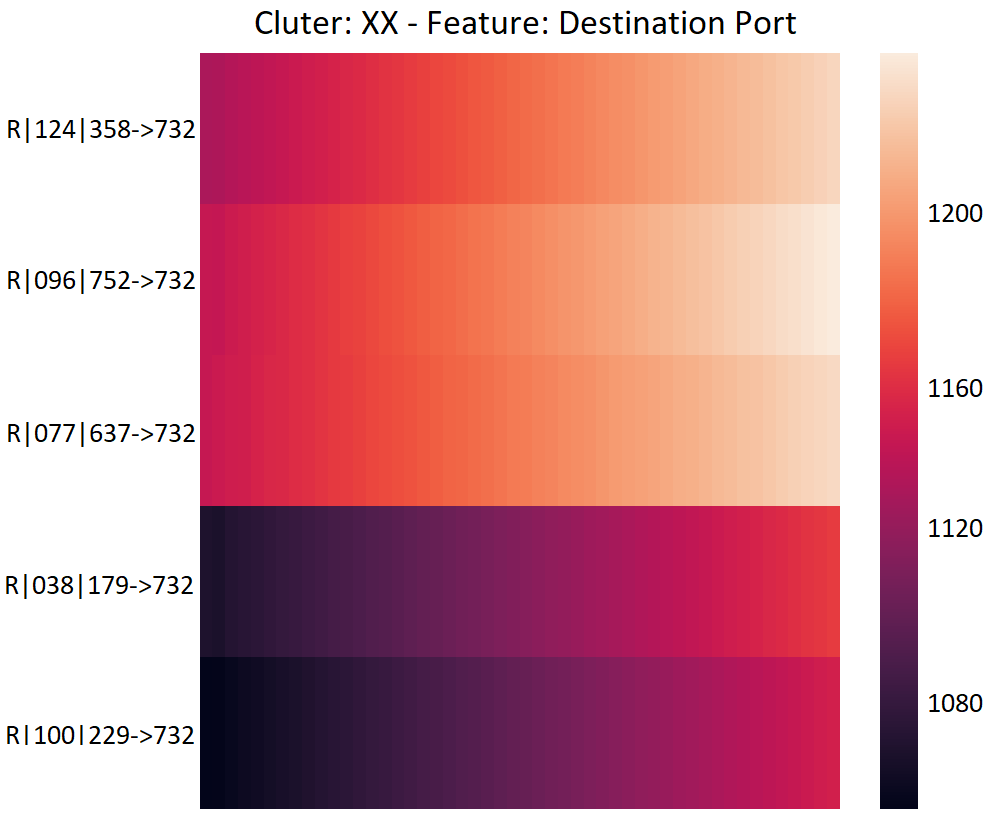}}}%
    \subfloat[Randomized port scan.]{{\includegraphics[width=0.5\linewidth]{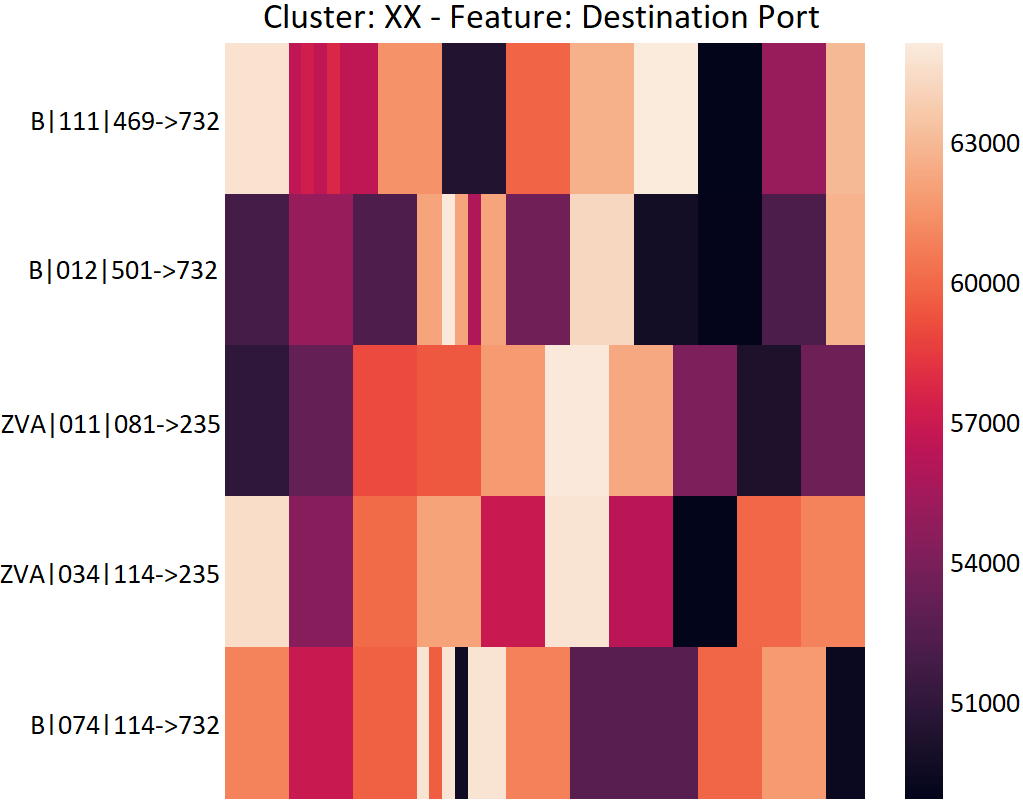}}}%
    \caption{Clusters showing systematic and randomized port scans}
    \label{fig:portscan}%
\end{figure}

\item\textbf{C\&C Reuse by Multiple Families.} One cluster contains connections from different families that contact the same C\&C server, and their temporal heatmaps look behaviorally identical. The cluster includes three \texttt{Zeus-Panda} (ZPA) connections and one \texttt{Blackmoon} (B) connection who contact a single IP address (encoded as 009), which has been reported as malicious. 
Figure \ref{fig:reuse} shows the temporal heatmaps of this cluster. The said connections are highlighted in green.
This result suggests that either the YARA rules mislabeled one of the samples or that authors share 
C\&C servers.

\begin{figure} [ht]
    \centering
    \subfloat[Packet sizes]{{\includegraphics[width=0.5\linewidth]{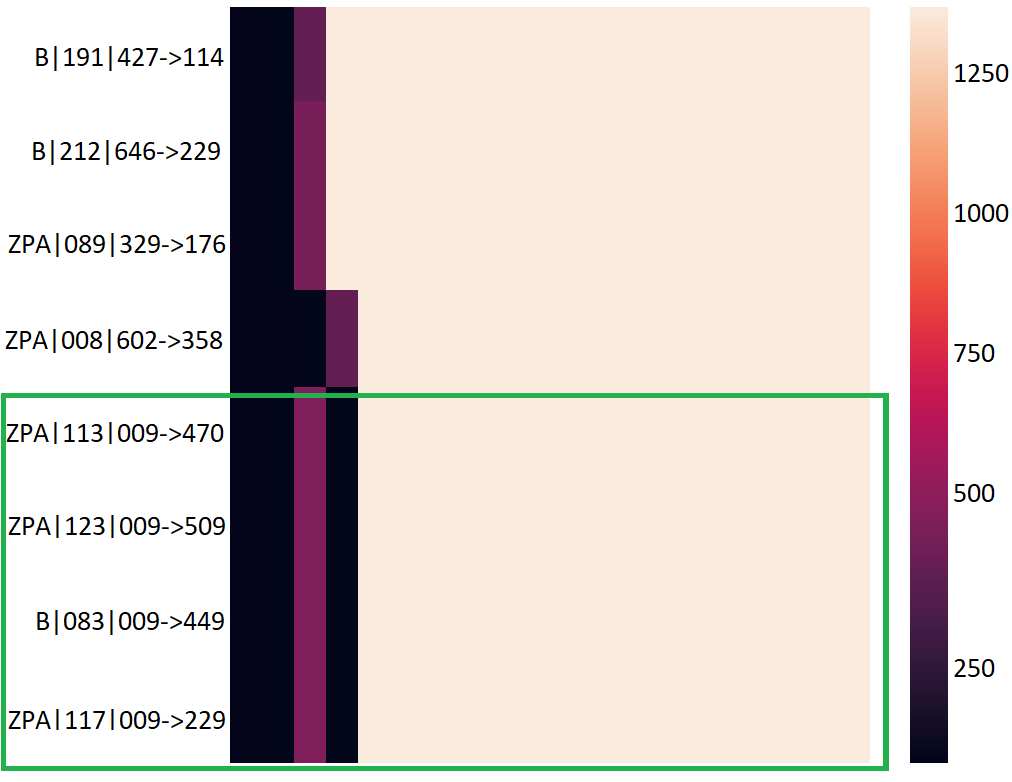}}}%
    \subfloat[Interval]{{\includegraphics[width=0.5\linewidth]{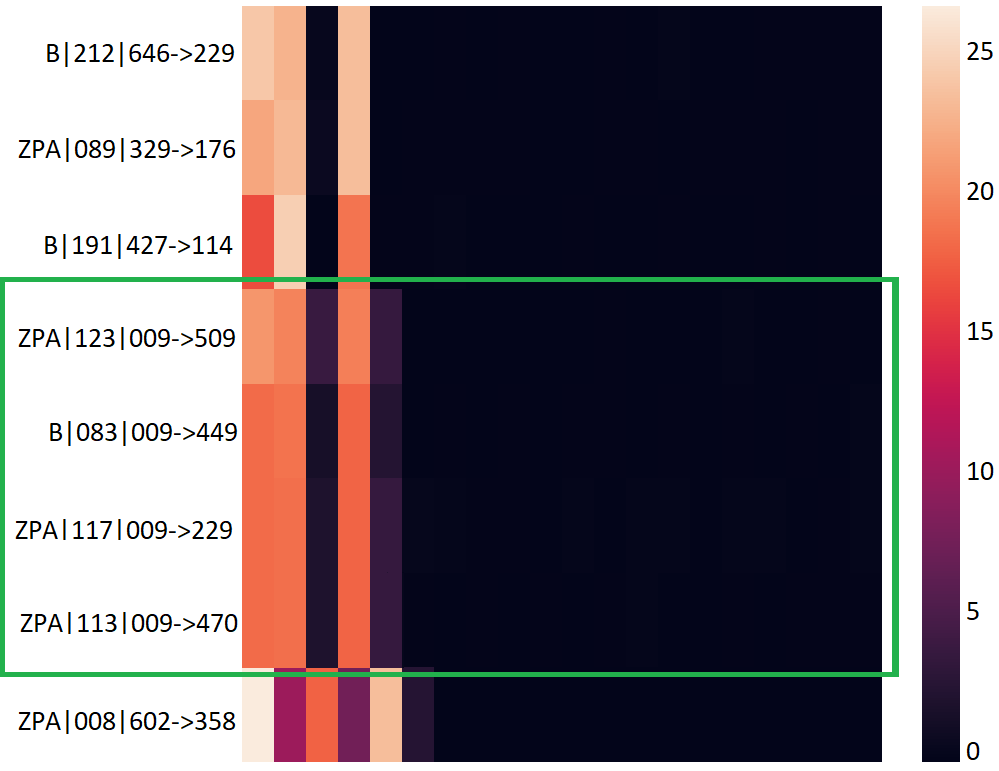}}}%
    \caption{Similar Zeus-Panda and Blackmoon connections.}
    \label{fig:reuse}%
\end{figure}

\item\textbf{Malicious Subnet Identification.} In some instances, several connections contact IP addresses that fall in the same subnet. For example,
two \texttt{Zeus-VM-AES} connections contact one host from 62.113.203.XX subnet, while 
another connection detected 15 days later contacts another host in the said subnet. Similarly, two \texttt{Zeus-Panda} connections and one Blackmoon connection contact two hosts in 88.221.14.XX subnet.
This gives actionable intelligence to ISPs to investigate if other IPs in these subnets are also hosting C\&C servers. 
%
\end{enumerate}

\subsection{\textbf{Cluster Characterization}}
We analyze the temporal heatmaps for the behavioral trend of each cluster in order to label it. MalPaCA's goal is to identify different behaviors in the network traffic and it does so regardless of their maliciousness and origin. Hence, the resulting clusters contain both, benign and malicious behaviors. The common clusters can be discarded if they contain known-benign behaviors, drastically reducing the number of connections to analyze. 

We successfully assigned labels to 12 clusters. For example, in the case of connection spam, the whole cluster is filled with almost identical connections originating from the same host. We validate this observation by specifically looking at the network traffic of these connections to see exactly what behavior is shown. Six clusters were left unlabeled since we could not identify the captured capability simply by exploring temporal heatmaps. These particular clusters were also the source of clustering errors. 
Table \ref{tab:rarecommon} shows that \textit{SSDP} and \textit{Broadcast traffic} are the most common behaviors and are both specific to Windows OS. Since the dataset is composed of Windows-based malware, it explains why 9 out of 12 families have connections in these two clusters. On the contrary, \textit{Connection Spam} and \textit{Malicious Subnet} are the rarest behaviors. \textit{Malicious Subnet} only captures \texttt{Zeus-VM-AES}. \texttt{Gozi-ISFB} opens numerous connections, creating a \textit{Connection Spam}. The incoming connections are stored in one cluster, while the outgoing traffic is split into two clusters due to the difference in the type of requests. This detailed behavioral analysis enables the identification of interesting clusters to analyze further. 

\runinhead{Performance Analysis.} The temporal heatmaps show that on average, 8.3\% connections per cluster are CEs---their feature sequences are different from their fellow connections in a cluster. 
The majority of the errors originate from the last six clusters. Note that this error rate is low for an unsupervised setting since not all connections require manual revision.

\subsection{\textbf{Constructing Behavioral profiles}}
MalPaCA identifies 18 distinct behaviors in the dataset. Hence, each malware sample (and its associated Pcap file) can be described as a binary string of 18 characters, known as \textit{Cluster Membership String (CMS)}, where each character signifies whether the Pcap's connections were found in that cluster. Precisely, for a malware sample $x$, $CMS_x = {b}^n$, where $b \in \{0,1\}$, $n$ is the number of behavioral clusters, and $b^i$ indicates whether $x$'s connections are present in the $i^{th}$ cluster.  
The Cluster Membership String can be regarded as the behavioral profile of a given malware sample. In this work, we consider binary CMSs because we are only interested in the behavior overlap of different malware samples. Non-binary $CMS_x = {z}^n$, for connection counts $z \in \mathbb{Z}$, is an interesting avenue to investigate.

Table \ref{tab:familybehaviors} lists the composite behavioral profiles for each YARA malware family in the dataset--- each YARA family is represented as the union of all its samples' CMSs.  \texttt{Dridex}, \texttt{Gozi-EQ}, \texttt{Zeus-P2P} and \texttt{Zeus-v1} only generate either \textit{SSDP} or \textit{Broadcast traffic}. Since this traffic is obtained from standard Windows services, it is likely that the malware was not activated when the associated Pcap files were recorded. Hence, the only connections observed from these families seem benign. 
\texttt{Gozi-ISFB} has the most diverse profile, with connection in 16 out of 18 clusters, which exhibit attacking capabilities such as \textit{Port Scans} and \textit{Connection Spamming}. Specifically, the \textit{Connection Spamming} behavior is never exhibited by any other malware family in the dataset. There are two reasons for \texttt{Gozi-ISFB}'s diversity: (i) \texttt{Gozi-ISFB} is the largest family under consideration, so many of its behavioral aspects are captured; and (ii) \texttt{Gozi-ISFB} opens more connections per sample compared to other families. For example, one sample of \texttt{Gozi-ISFB} opens 111 connections, while the average number of connections for other malware samples is 3.

\begin{table*}[ht]
\caption{Composite behavioral profiles of malware families. Columns: YARA labels, Rows: Cluster labels by MalPaCA.}
\label{tab:familybehaviors}
\centering
\begin{tabular}{l|c|c|c|c|c|c|c|c|c|c|c|c}
\toprule
\multicolumn{1}{c}\textbf{   } & \multicolumn{1}{c}{\textbf{   B  }} & \multicolumn{1}{c}{\textbf{   C}} & \multicolumn{1}{c}{\textbf{   D  }} & \multicolumn{1}{c}{\textbf{  DL }} & \multicolumn{1}{c}{\textbf{  GE }} & \multicolumn{1}{c}{\textbf{  GI }} & \multicolumn{1}{c}{\textbf{   R  }} & \multicolumn{1}{c}{\textbf{   Z  }} & \multicolumn{1}{c}{\textbf{  ZP }} & \multicolumn{1}{c}{\textbf{ ZPa}} & \multicolumn{1}{c}{\textbf{ Zv1}} & \multicolumn{1}{c}{\textbf{ ZVA}} \\ \midrule
\textbf{\begin{tabular}[c]{@{}l@{}}SSDP  traffic\end{tabular}} & \OK & \OK & \OK & \OK & \OK & \OK & \OK & \OK & - & \OK & - & \OK \\ 
\textbf{\begin{tabular}[c]{@{}l@{}}Broadcast  traffic\end{tabular}} & \OK & \OK & - & \OK & - & \OK & \OK & - & \OK & - & \OK & \OK \\ 
\textbf{\begin{tabular}[c]{@{}l@{}}LLMNR  traffic\end{tabular}} & \OK & \OK & - & \OK & - & \OK & - & - & - & - & - & - \\ 
\textbf{\begin{tabular}[c]{@{}l@{}}System. port scan\end{tabular}} & \OK & \OK & - & - & - & \OK & \OK & - & - & - & - & \OK \\ 
\textbf{\begin{tabular}[c]{@{}l@{}}Random.  port scan\end{tabular}} & \OK & \OK & - & - & - & \OK & \OK & - & - & - & - & \OK \\ 
\textbf{\begin{tabular}[c]{@{}l@{}}In conn  spam\end{tabular}} & - & - & - & - & - & \OK & - & - & - & - & - & - \\ 
\textbf{\begin{tabular}[c]{@{}l@{}}Out conn  spam\end{tabular}} & - & - & - & - & - & \OK & - & - & - & - & - & - \\ 
\textbf{\begin{tabular}[c]{@{}l@{}}Malicious  Subnet\end{tabular}} & - & - & - & - & - & - & - & - & - & - & - & \OK \\ 
\textbf{\begin{tabular}[c]{@{}l@{}}In  HTTPs\end{tabular}} & - & \OK & - & \OK & - & \OK & - & - & - & \OK & - & - \\ 
\textbf{\begin{tabular}[c]{@{}l@{}}Out  HTTPs\end{tabular}} & - & - & - & - & - & \OK & - & - & - & \OK & - & - \\ 
\textbf{\begin{tabular}[c]{@{}l@{}}C\&C  reuse\end{tabular}} & \OK & - & - & - & - & - & - & - & - & \OK & - & - \\ 
\textbf{Misc.} & \OK & \OK & - & \OK & - & \OK & - & \OK & - & \OK & - & \OK \\ \midrule \midrule
\textbf{\begin{tabular}[c]{@{}l@{}} \# Clusters\end{tabular}} & 7 & 11 & \textit{\textbf{1}} & 8 & \textit{\textbf{1}} & \textit{\textbf{16}} & 4 & 2 & \textit{\textbf{1}} & 7 & \textit{\textbf{1}} & 7 \\ \bottomrule
\end{tabular}
\end{table*}

\subsection{\textbf{Showing Relationships using DAG}}
We extract the behavioral relationships between the 216 Cluster Membership Strings by considering it a \textit{Set Membership} problem. It dictates that, e.g. \texttt{Set A= \{0,1,1\}} is a \textit{subset} of \texttt{Set B=\{1,1,1\}} because \texttt{Set B} encapsulates all of \texttt{Set A}'s behaviors and more. Similarly, \texttt{Set C= \{0,0,0\}} is a subset of every other set in this domain. \texttt{Set C} represents Pcaps where all connections were discarded as Noise due to significant differences in behavior. 

\begin{figure*}[ht]
    \includegraphics[width=\linewidth]{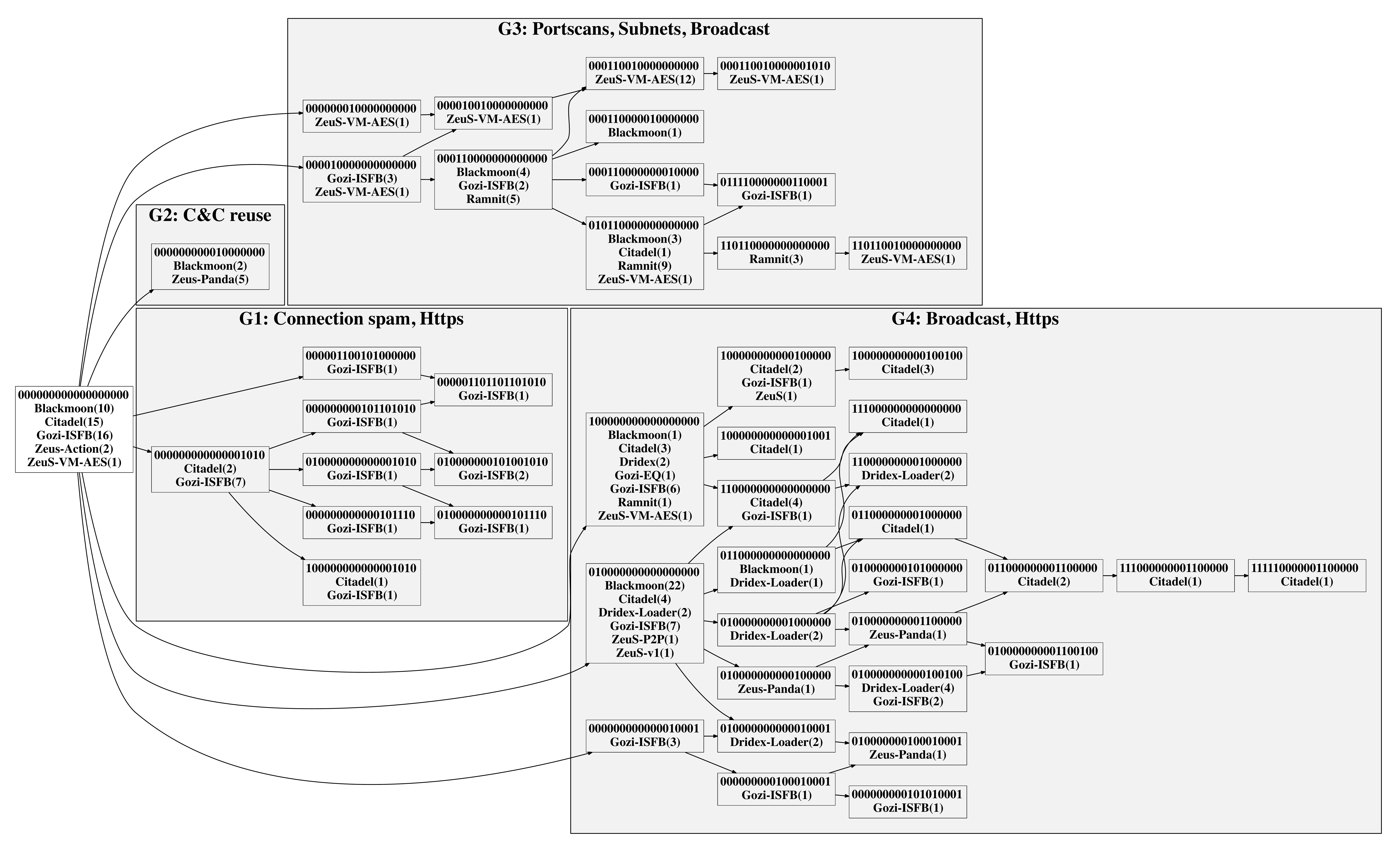}
    \caption{MalPaCA's behavioral profiles: The DAG shows the behavioral relationships between malware samples. Each node shows a CMS, and compares with YARA family labels (+ \# Pcaps).}
    \label{fig:coloreddag}
\end{figure*}

We represent the relationships between Pcap files using a Directed Acyclic Graph (DAG), shown in Figure \ref{fig:coloreddag}. Each node represents a unique Cluster Membership String. Multiple Pcaps can share a single CMS IFF their behaviors overlap. The nodes with minimum Hamming distance are connected using edges. This method allows multiple parents, i.e. a CMS of \texttt{"111"} may be reached by both \texttt{"110"} and \texttt{"101"}. Note that this graph is constructed purely from a data-driven approach without using any knowledge of family labels. In combination with human intelligence, we believe that it can serve as a powerful tool in understanding malware's network behavior.

\section{Comparative analysis}
We show MalPaCA's results in relation to existing work by conducting two comparative analyses: (i) Comparing MalPaCA's behavioral profiles with YARA family labels, and (ii) Comparing MalPaCA's cluster quality with an existing approach that uses statistical features.  

\subsection{Comparison with Traditional Family labels} \label{sec:comparison-av-labels}

We use the DAG from Figure \ref{fig:coloreddag} to contrast between YARA labels and MalPaCA's behavioral profiles. 
Each node shows a unique CMS, and the number of malware families that share it. For example, the node with the CMS of \texttt{"000000000000001010"} is labeled as \texttt{"Citadel(2), Gozi-ISFB(7)"} because 2 \texttt{Citadel} Pcaps and 7 \texttt{Gozi-ISFB} Pcaps show the same behavior --- their connections are co-located in the clusters 15 and 17. The root (on the left most side) contains the Pcaps for which all connections were discarded as Noise. Pcaps showing subsequently more behaviors are placed towards the right of the graph, with the right most node \texttt{"111110000001100000 Citadel(1)"} containing one \texttt{Citadel} Pcap that shows the most diverse number of behaviors. Note that observing additional network traffic will enrich this graph even further. 

The graph shows four major partitions (denoted by G1-G4), indicating that there are four high-level behavioral sub-groups present in the dataset. 
The G2 group containing only one node stands out. It contains Pcaps from \texttt{Zeus-Panda} and \texttt{Blackmoon}, and are the only malware samples that share a C\&C server. This observation makes a strong case that these particular Pcap files, albeit originating from two families, are behaviorally alike.
The G3 group contains Pcaps from various families that are observed doing port scans and broadcasting behaviors. Some servers from this group also form malicious subnets. The G4 group, on the other hand, is the largest group that uses HTTPs traffic along with broadcasting behaviors. The G1 group is highly dominated by \texttt{Gozi-ISFB} and is observed doing Connection spamming, along with using HTTPs traffic. Some connections from these \texttt{Gozi-ISFB} Pcaps were placed in the behavioral clusters that we failed to identify (c13-c18).   

The node location for some malware families is intriguing. For example, most of the \texttt{Zeus-VM-AES} Pcaps that are associated with malicious subnets are located in the G3 group, together with \texttt{Ramnit} files that are associated with port scans. 
\texttt{Dridex-Loader} is only observed in group G4, while most of the \texttt{Citadel} Pcaps are also seen in the same. 
\texttt{Blackmoon} and \texttt{Gozi-ISFB} have Pcaps that are distributed over all of the behavioral sub-groups. However, \texttt{Gozi-ISFB} is seen dominating the G1 group, while \texttt{Backmoon} dominates the G4 group. Furthermore, as observed from Table \ref{tab:familybehaviors}, \texttt{Gozi-ISFB}'s Pcaps collectively show 18 discrete behaviors and \texttt{Citadel}'s Pcaps show 11 behaviors. However, \texttt{Citadel} shows more discrete behaviors in a single Pcap compared to \texttt{Gozi-ISFB}, as \texttt{Gozi-ISFB}'s Pcaps contain more (behaviorally similar) connections on average. Also, each of \texttt{Gozi-ISFB}'s Pcaps is more behaviorally dissimilar than \texttt{Citadel}'s Pcaps.    

\texttt{Zeus-Panda}'s Pcaps are clearly divided in two behavioral sub-groups --- one in G2 group with \texttt{Blackmoon} samples and the other in the G4 group. \texttt{Zeus-v1}, \texttt{Zeus-P2P}, \texttt{ZeuS}, \texttt{Gozi-EQ}, and \texttt{Dridex} are only seen at the left side of the graph, indicating that none of their distinguishing behaviors were present in the dataset.

To conclude, the DAG clearly identifies the discrepancies in the malware's behavioral profiles and their traditional family names. A significant portion of the analysis pipeline is automated and unsupervised. The temporal heatmaps together with the DAG are intended for human-in-the-loop exploration --- they actively support malware behavior analysis and provide more insightful characterization of malware than current family labels.

\subsection{\textbf{Comparison with Statistical Features}}

\runinhead{Baseline Set-up.} 
We compare the cluster quality of using sequential versus statistical features.
We use the existing method by Tegeler et al.~\cite{botfinder} (called baseline, henceforth) to compare our results since they not only use statistical features, but also incorporate periodic behavior using Fourier transform to detect bot-infected network traffic. 
Although the goal of their study diverges from ours, their feature selection approach is aligned with ours. For objectivity, we keep the rest of the pipeline as explained in Section \ref{sec:method}. 
Taking guidelines from Tegeler et al.~\cite{botfinder} and adapting them to our problem statement, each connection in the baseline is characterized by 1) average packet size, 2) average interval between packets, 3) average duration of a connection and 4) the maximum Power Spectral Density (PSD) of the FFT obtained by the binary sampling 
approach by Tegeler et al.~\cite{botfinder} -- the signal is 1 when a packet is present in the connection and is 0 in between.

\runinhead{Cluster quality comparison.} The baseline method results in 22 clusters, with an average of 21.2 connections per cluster. 265 connections are discarded as noise. These results are in comparison with sequence clustering -- 18 clusters; on average 25 connections per cluster; 284 connections discarded as noise. 

Baseline seems to perform better with smaller cluster size on average and discarding fewer connections as noise. However, a deeper analysis shows the obtained clusters lack quality. 

\begin{figure} [t]
    \centering
    \subfloat[Packet sizes]{{\includegraphics[width=0.5\linewidth]{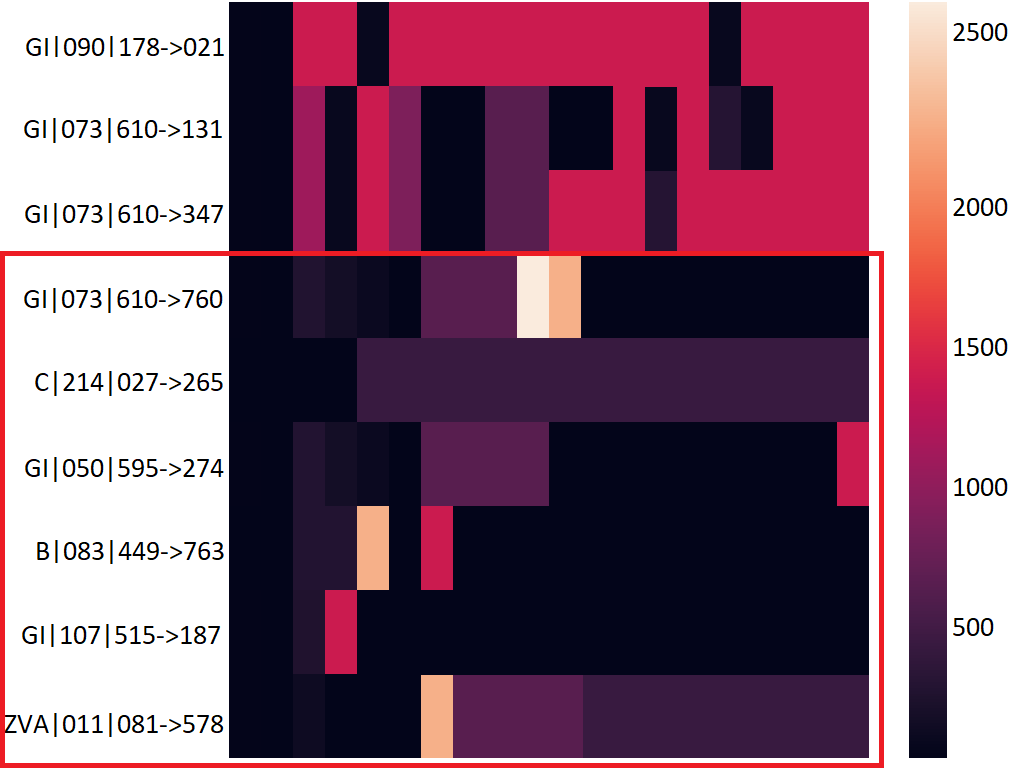}}}%
    \subfloat[Interval]{{\includegraphics[width=0.5\linewidth]{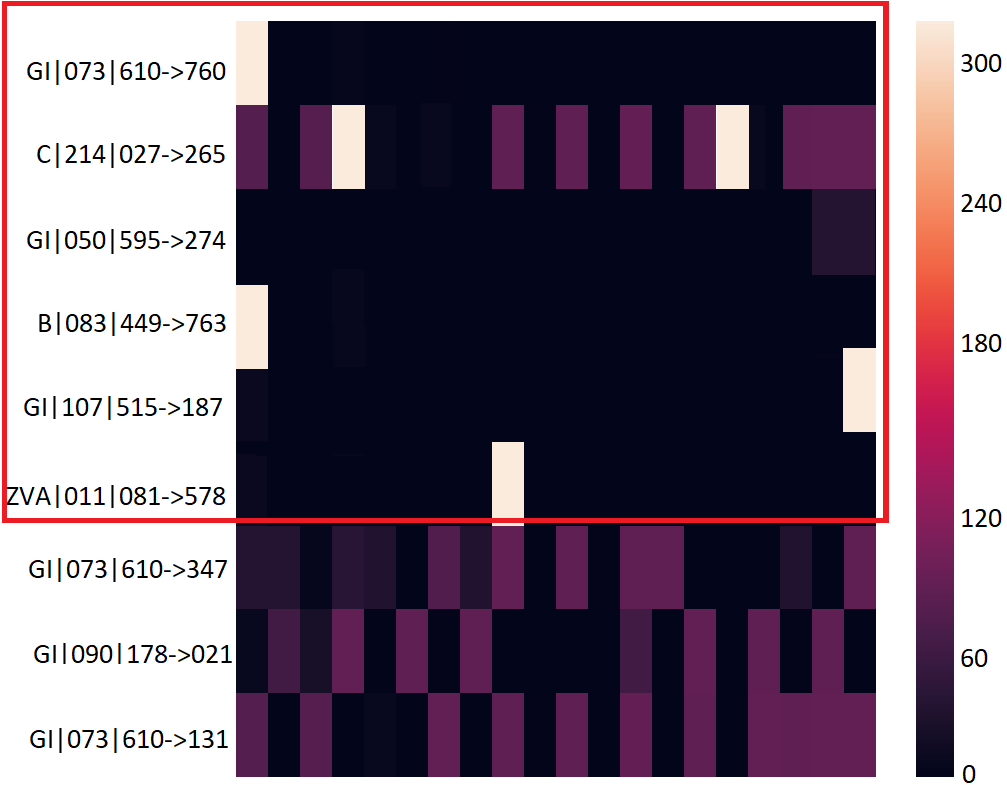}}}%
    \caption{Baseline clusters: Six out of nine behaviorally different connections clustered together in baseline version.}%
    \label{fig:blmisplaced}%
\end{figure}

\begin{enumerate}
    \item With statistical features, connections present in most clusters appear very different from their fellow connections. On average, 57.5\% connections per cluster have visually different temporal heatmaps, compared to 8.3\% for sequential features. Figure \ref{fig:blmisplaced} shows a cluster from the baseline. It has nine connections, out of which six are errors based on their behavior. The \textit{rightful owners} of the cluster are the connections that have the least mutual distance, i.e. \texttt{GI|090|178$\rightarrow$021}, \texttt{GI|073|610$\rightarrow$131}, \texttt{GI|073|610$\rightarrow$346}. The other six connections have minor differences in all features, except the source port which is 6 for all. They were clustered together because their statistical features had the least mutual distance, i.e. $average\_time\_interval\allowbreak = 19.77 \pm3.11$; $fft = 0.07 \pm 0.05$; $average\_duration = 397.7 \pm 61.7$; $\allowbreak average\_bytes\allowbreak = 573.3 \pm 113.8$.
The temporal heatmaps clearly show behavioral differences in nearly all clusters.
\item Statistical features are also unable to identify the direction of network traffic. In the cluster shown in Figure \ref{fig:blmisplaced}, there is one incoming connection in the cluster along with eight outgoing ones. A similar trend is observed for 19 out of 22 clusters. In contrast, sequences of packet size and inter-arrival time are enough to identify traffic direction in sequence clustering.
\end{enumerate}

In summary, while statistical features may be simple to use, they lose behavioral information that plays a crucial role in accurately determining similarities in network behavior. 
Sequence clustering obtains significantly better clusters. 
Given that modeling behavioral profiles is already challenging for short sequences, it is remarkable that MalPaCA can identify network behaviors using only 20 packets and 4 coarse features.

%% file: Sec-Limitations.tex
\section{\textbf{Limitations and Future work}}\label{sec:lim}

\runinhead{Limitations.} 
Performance optimizations are needed to make sequence clustering more efficient and scalable. 
In MalPaCA, DTW forms the main bottleneck as the length of sequences grows longer. There exist streaming versions of DTW that compute results in real-time. One such technique is presented by Oregi et al. \cite{oregi2017onlinedtw}. Moreover,
using Locality Sensitive Hashing \cite{bayer2009scalable, azab2014mining} can make MalPaCA more scalable.

Density-based clustering discards rare events as noise. This makes sense if the dataset is noisy. However, in the presence of a purely malicious dataset, the connections that lie in lower-density regions may represent rare attacking capabilities, which may be discarded in the current implementation. 

Malware authors can try to evade detection by modifying malware's code. A common assumption is that malware can easily evade detection by adding random delays and padding to packets. However, there is a limit to what an attacker can change. For example, a TCP handshake needs to happen in a certain way because this is how the protocol dictates it. Also, padding-related provisions are already standardized by some commonly used protocols, such as TLS making it 
difficult to hide ``coarse" features like packet sizes and inter-arrival times \cite{dyer2012peek}. We expect that MalPaCA is evasion resilient, e.g. since MalPaCA only uses coarse features, evading it is not a trivial task. Moreover, the usage of Dynamic Time Warping distance makes it resilient to random delays \cite{elfeky2005dtwnoiseresilient} and due to the relative distance measures used in HDBScan, randomized port numbers are already clustered together, as shown in Section \ref{sec:results}. If, after all this, attackers still manage to evade MalPaCA, the malware sample will end up with a new behavioral profile, making analysts more prone to analyze it. More study is needed to strengthen these claims.

\runinhead{Future work.} 
There are several research directions this work can take: (i) We will work on fully automating the capability assessment of malware by building a directory of observed behaviors, which will be used for cluster labeling. (ii) We will test and improve MalPaCA's adversarial evasion resilience. (iii) We will integrate additional behavioral data sources in MalPaCA so the profiles are based on all static, system-level and network behavior. (iv) Since MalPaCA is a generic technique, we will test its applicability in building behavioral profiles for everyday-use software.

%% file: Sec-Conclusions.tex
\section{Conclusions}

In this chapter, we propose MalPaCA, an intuitive network traffic-based tool to perform malware capability assessment: It groups capabilities using sequence clustering, and uses the cluster membership to build network behavioral profiles. 
We also propose a visualization-based cluster evaluation method whose key advantage is its white-box nature, allowing malware analysts to investigate, understand, and even correct labels, if necessary. 
We implement MalPaCA and evaluate it on real-world financial malware samples collected in the wild. 
MalPaCA independently identifies attacking capabilities. We build a DAG to show overlapping malware behaviors, and discover a number of samples that do not adhere to their family names, either because of incorrect labeling by black-box solutions or extensive overlap in the families' behavior.
We also show that sequence clustering outperforms existing statistical features-based methods by making only 8.3\% errors, as opposed to 57.5\%.
MalPaCA, with its visualizations and capability assessment, can actively support the understanding of malware samples. 
The resulting behavioral profiles give malware researchers a more informative and actionable characterization of malware than current family designations.